\input{epsf}

\newcommand{\BOX}{\hbox {$\sqcap$ \kern -1em $\sqcup$}}

\newcommand{\hf}{{1\over 2}}

\newcommand{\R}{{\Bbb R}}
\newcommand{\C}{{\Bbb C}}

\renewcommand{\to}{\rightarrow}
\newcommand{\tensor}{\otimes}
\newcommand{\maps}{\colon}

\newcommand{\F}{{\cal F}}
\newcommand{\E}{{\cal E}}
\newcommand{\V}{{\cal V}}

\newcommand{\SU}{{\rm SU}}

        \newcommand{\be}{\begin{equation}}
        \newcommand{\ee}{\end{equation}}
        \newcommand{\ba}{\begin{eqnarray}}
        \newcommand{\ea}{\end{eqnarray}}
        \newcommand{\ban}{\begin{eqnarray*}}
        \newcommand{\ean}{\end{eqnarray*}}
        \newcommand{\barr}{\begin{array}}
        \newcommand{\earr}{\end{array}}

\newcounter{letter} \newcounter{numeral} \newcounter{Numeral}

\newenvironment{Alphalist}{
\begin{list}{(\Alph{letter})}{\usecounter{letter}}
}{\end{list}}

\documentstyle[12pt,amsfonts]{article}

\begin{document}

      \begin{center}
      {\bf Spin Foam Perturbation Theory\\}
      \vspace{0.5cm}
      {\em John C.\ Baez\\}
      \vspace{0.3cm}
      {\small Department of Mathematics, University of California\\ 
      Riverside, California 92521 \\
      USA\\ }
      \vspace{0.3cm}
      {\small email: baez@math.ucr.edu\\}
      \vspace{0.3cm}
      {\small October 13, 1999\\ }
      \end{center}

\begin{abstract}   
\noindent    
We study perturbation theory for spin foam models on triangulated
manifolds.  Starting with any model of this sort, we consider an
arbitrary perturbation of the vertex amplitudes, and write the 
evolution operators of the perturbed model as convergent power series in
the coupling constant governing the perturbation.   The terms in the
power series can be efficiently computed when the unperturbed model is
a topological quantum field theory.  Moreover, in this case we can
explicitly sum the whole power series in the limit where the number 
of top-dimensional simplices goes to infinity while the coupling constant
is suitably renormalized.  This `dilute gas limit' gives spin 
foam models that are triangulation-independent but not topological
quantum field theories.  However, we show that models of this sort
are rather trivial except in dimension 2.
\end{abstract}

\section{Introduction}

Recent work on loop quantum gravity and topological quantum field theory
has focused attention on a class of theories called `spin foam models'
\cite{B2}.  In a spin foam model, states are described as linear
combinations of spin networks, while transition amplitudes are computed
as sums over spin foams.  A `spin network' is a graph with edges
labelled by representations of some group or quantum group, and with
vertices labelled by intertwiners.  Similarly, a `spin foam' is a
2-dimensional complex with polygonal faces labelled by representations
and edges labelled by intertwiners.   For each spin foam one calculates
an amplitude as a product of amplitudes associated to its faces, edges
and vertices.   The transition amplitude between spin network states 
is computed by summing the amplitudes of spin foams that go from one
spin network to another.  

There are various versions of this basic idea. The most radical and 
perhaps ultimately most promising option is to work with `abstract' spin
networks and spin foams, not embedded in any background manifold
\cite{B1,DFKR,MS,S2}.  A more conservative approach is to describe states by
spin networks embedded in a manifold that represents space, and
similarly, to sum over spin foams embedded in a manifold that represents
spacetime.   This strategy has the advantage of making more rapid
contact with familiar physics.  However, it has certain problems that do
not arise --- or at least, not so obviously --- in the `abstract'
approach.  In this paper we consider one of these, namely, the problem
of triangulation-dependence.  

When summing over spin foams embedded in a given spacetime manifold, 
one has a number of choices.  One could try to sum, or integrate, over
{\it all} such spin foams.  Unfortunately the set of all such spin
foams is uncountable, and there is no obvious measure on it.  A somewhat
more promising alternative is to sum over all diffeomorphism equivalence
classes of embedded spin foams.   This idea is rather natural if we
start with a formula for the Hamiltonian constraint in canonical quantum
gravity and try to derive a spin foam model from that \cite{RR,Rovelli2}.
This set is still uncountable if we work with spin foams smoothly
embedded in a smooth manifold, since there are moduli spaces of
intersections \cite{Rovelli}, but it becomes countable in the 
piecewise-linear context.  Of course, even with a countable sum one
faces the issue of convergence.  

A more drastic way to obtain well-defined transition amplitudes is to
fix a triangulation of the spacetime and sum only over spin foams that
live in the dual 2-skeleton of this triangulation.  This approach has 
been very successful for models that give topological quantum 
field theories, such as:
\begin{itemize}
\item The Fukuma-Hosano-Kawai version of the $G/G$ gauged
WZW model in 2 dimensions, where $G$ is a compact Lie group \cite{FHK}. 
\item The Turaev-Viro model of 3-dimensional $\SU(2)$ 
$BF$ theory with cosmological term, and its subsequent generalization to other 
compact Lie groups \cite{BW,DJN,Turaev,TV}. 
\item The Crane-Yetter model for 4-dimensional $\SU(2)$
$BF$ theory with cosmological term, and its subsequent generalization
to other compact Lie groups \cite{CKY,CY}.  
\item The Dijkgraaf-Witten model, and its generalization to spacetimes
of arbitrary dimension \cite{DW,FQ,P,Y}.  Here the gauge group is a
finite group.
\end{itemize}
The reason is that in these cases, the sum over spin foams is
finite and the result is independent of the choice of triangulation.
This allows us to have our cake and eat it too: we obtain the
benefits of finiteness that come from working with a triangulation,
while still being able to regard the triangulation as a mere
computational device, rather than something `physical'.  

However, in the existing spin foam models of 4-dimensional quantum
gravity that involve fixing a triangulation of spacetime
\cite{BC,BC2,Iwasaki,R}, the results appear to depend on the
triangulation.  There are various attitudes one can take to this.  One
can accept the triangulation as an inevitable fixed background 
structure.  However, this goes against the desire for a background-free
theory of quantum gravity.   Alternatively, one can try to eliminate the
triangulation-dependence somehow: for example, either by summing over
triangulations, or by taking a limit as the triangulation becomes ever
finer.  Both these options lead to questions of convergence.  

Since spin foam models corresponding to topological quantum field
theories are well-understood, while others are not, it seems natural to
develop a perturbation theory for spin foam models.  This might allow 
us to study nontopological theories as perturbations of topological 
ones.  This idea is already present in the work of Freidel
and Krasnov \cite{FK}, who consider a large class of spin foam models as
perturbations of $BF$ theory.   It is also related to Smolin's work
on strings as perturbations of evolving spin networks \cite{S1}.

In this paper we start with any spin foam
model that involves a fixed triangulation of a manifold representing
spacetime, and consider an arbtrary perturbation of the vertex
amplitudes of this model.   (Perturbing the edge and face amplitudes
works similarly, so to reduce the complexity of our notation we do
not consider such perturbations.)  We expand the evolution operators 
of the perturbed model as convergent power series in the coupling constant
governing the perturbation.    

The terms in this power series can be efficiently computed when the 
unperturbed model is a topological quantum field theory, such as those 
listed above.  In fact, in this case the power series can be explicitly 
summed in the limit where the number of top-dimensional simplices in 
the triangulation goes to infinity, as long as we also renormalize 
the coupling constant.   This limit corresponds to a `dilute gas of 
interactions'.   Taking this limit, we obtain models that are
triangulation-independent but not topological quantum field
theories.  Examples include 2d Yang-Mills theory and a host of other
theories in 2 dimensions.  However, we show that the dilute gas limit
does not give interesting spin foam models in higher dimensions.  We
discuss the implications of this fact in the Conclusions.

In what follows, we assume some familiarity with the basic definitions
concerning spin networks and spin foams.   These can be found in the
references \cite{B0,B1,B2,S0}.   

\section{Partition Functions}

In this section we give a power series for the partition function of a
spin foam model obtained by perturbing the vertex amplitudes of a fixed
`unperturbed' model.  For simplicity we restrict attention to the case
when spacetime is a compact manifold without boundary.  In the next
section we generalize this result to the case when spacetime is compact
manifold with boundary (i.e., a cobordism).  

Let $M$ be a compact connected oriented piecewise-linear $n$-manifold
equipped with a triangulation $\Delta$.  Let $\V$ (resp.\ $\E$, $\F$) be
the set of 0-cells (resp.\ 1-cells, 2-cells) of the dual 2-skeleton of
this triangulation.  We call these `vertices', `edges', and `faces',
respectively.  We assume that  each face and edge is equipped with an
orientation.   By Poincar\'e duality, we may also think of $\V$ (resp.\
$\E$, $\F$) as the set of simplices in $\Delta$ with dimension $n$
(resp.\ $n-1$, $n-2$).   We find it useful to freely switch between
these points of view.  

We now fix a spin foam model to perturb about.  We shall not need much
detailed information about this model.  First we fix a suitable category
--- typically a category of representations of some group or quantum
group.  We assume that this category is equipped with a tensor product,
and that every representation in the category is a direct sum of
representations chosen from some finite set $R$ of inequivalent
irreducible representations.    Finally, we assume that the hom-sets of
our category are finite-dimensional Hilbert spaces.   The topological
quantum field theories listed in the Introduction satisfy all these 
assumptions.

Given this, a `spin foam' is a way of labelling each face in $\F$ with a
representation in $R$ and labelling each edge in $\E$ with an
intertwiner chosen from a fixed orthonormal basis of $\hom(\rho_1
\tensor \cdots \tensor \rho_n, \rho'_1 \tensor \cdot \tensor \rho'_m)$,
where $\rho_i$ are the representations labelling the faces incoming to
this edge, and $\rho'_j$ are the representations labelling faces
outgoing to this edge.   

We define the partition function of our unperturbed spin foam model 
by: 
\[    Z_0(M) = \sum_F \prod_{f \in \F} A_0(F,f) 
                      \prod_{e \in \E} A_0(F,e)
                      \prod_{v \in \V} A_0(F,v).  \]
Here the sum is taken over all spin foams $F$.  The `face amplitude'
$A_0(F,f)$ is a complex number depending on the representation 
that $F$ assigns to the face $f$.  The `edge amplitude' $A_0(F,e)$ 
is a complex number depending on the intertwiner that $F$ assigns
to the edge $e$.   Finally, the `vertex amplitude' $A_0(F,v)$ is a
complex number depending on the representations and intertwiners
labelling faces and edges incident to $v$.  

Next we perturb the vertex amplitudes of this spin foam model:
\ba    A(F,f) &=& A_0(F,f)  \label{perturb} \\
   A(F,e) &=& A_0(F,e)    \nonumber \\
   A(F,v) &=& A_0(F,v) + \lambda A_1(F,v)  . \nonumber \ea
The partition function of the perturbed model is given by:
\[    Z(M) = \sum_F   \prod_{f \in \F} A(F,f)  
                      \prod_{e \in \E} A(F,e)
                      \prod_{v \in \V} A(F,v) . \]
We may expand this partition function as a power series in the 
coupling constant $\lambda$:
\[
      Z(M) = \sum_{j = 0}^\infty \lambda^j Z_j(M) 
\]
where 
\[    Z_j(M) = \sum_{S \subseteq \V, \; |C| = j}  \sum_F 
                      \prod_{f \in \F} A(F,f)  
                      \prod_{e \in \E} A(F,e)
                      \prod_{v \notin C} A_0(F,v) 
                      \prod_{v \in C} A_1(F,v)  \]
In other words, the $j$th term in the perturbation expansion for the
partition function is a sum over spin foams where nontrivial
interactions occur at exactly $j$ vertices: the vertices in $C$.  We
call elements of $C$ `interaction vertices'.   The zeroth-order term,
$Z_0(M)$, is just the partition function of the original unperturbed 
spin foam model.  Note that this perturbation expansion converges for a
very simple reason: it is a finite sum!  The reason is that $Z_j(M) = 0$
when $j$ exceeds the total number of vertices in the dual 2-skeleton of
the triangulation of $M$.  We denote this number by $|\V|$.

To write the formula for $Z_j(M)$ more tersely, let us define a
`configuration' to be a triangulation $\Delta$ of $M$ together with a
given subset $C \subseteq \V$.   (We may also think of $C$ as a subset
of the $n$-simplices in the triangulation.)  For any configuration
$(\Delta,C)$, define 
\[    Z(\Delta,C) =  \sum_F \prod_{f \in \F} A(F,f)  
                      \prod_{e \in \E} A(F,e)
                      \prod_{v \notin C} A_0(F,v) 
                      \prod_{v \in C} A_1(F,v)       \]
where we sum over spin foams $F$ living in the dual 2-skeleton of the
triangulation.   Then we have 
\be Z_j(M) = \sum_{C \subseteq \V, \; |C| = j} Z(\Delta,C) . 
\label{zj2} \ee 

To make further progress, we need to know a bit more about our original
unperturbed spin foam model.  Let us assume that in the unperturbed
model, the partition function is independent of the triangulation of
$M$, including the choice of orientations for faces and edges.  More
importantly, let us assume that this can shown using purely {\it local}
calculations, by checking invariance under the Pachner moves
\cite{Pachner}.  All the topological quantum field theories listed in
the Introduction meet this assumption.  

Define two configurations to be `equivalent' if they become
combinatorially equivalent after repeatedly applying Pachner moves to
$n$-simplices that do not lie in the given subset.     Figure
\ref{configs} shows two equivalent configurations with the same
underlying triangulation.  In this example the manifold $M$ is the
2-sphere, and the 2-simplices in $C$ are colored black.  

\vbox{
\vskip 1em
\centerline{\epsfysize=2.0in\epsfbox{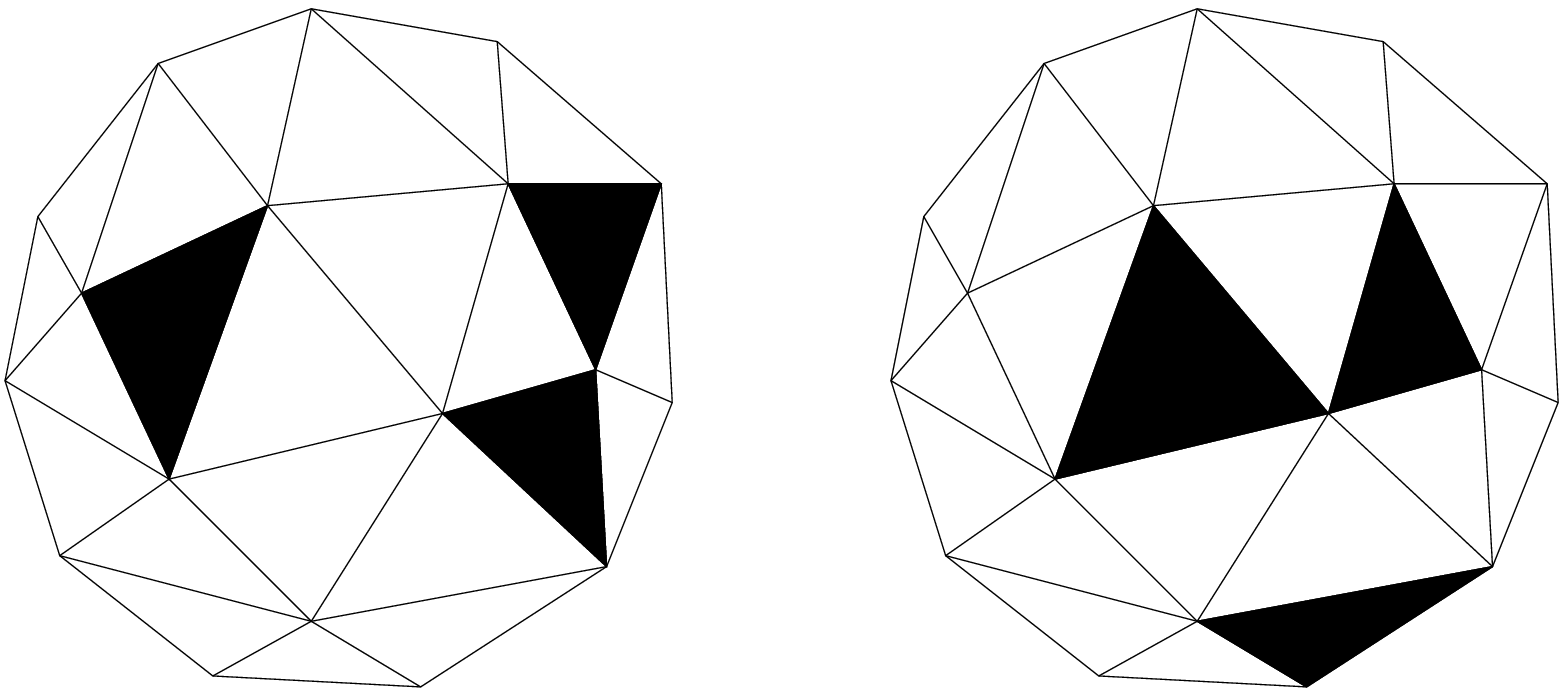}} \medskip
\centerline{1.  Two equivalent configurations with the same underlying
triangulation}
\label{configs}
\medskip
}

Define a `$j$-element configuration' to be one for which $|C| = j$.  
By our assumptions on the unperturbed spin foam model, $Z(\Delta,C)$ is
the same for any two configurations in the same equivalence class.   
This allows us to rewrite the sum in equation (\ref{zj2}) as a sum over
equivalence classes of $j$-element configurations $(\Delta,C)$, 
weighted by how many configurations are in each equivalence class.   

If we fix a triangulation $\Delta$, the number of $j$-element
configurations $(\Delta,C)$ is ${|\V| \choose j}$.  This is always
less than $|\V|^j / j!$, so keeping only the low-order terms in
the perturbation expansion for the partition function of the 
perturbed spin foam model should give a good approximation when 
\[         |\lambda|  \ll {1\over |\V|}  . \]
When $j$ is small, there should not be many equivalence classes of
$j$-element configurations.  We thus have a practical recipe for
approximately computing the partition function in this case.  

\section{Evolution Operators}

Now we turn to the evolution operators associated to cobordisms.   Our
unperturbed spin foam model assigns a Hilbert space $Z(S)$ of
`kinematical states' to any compact oriented $(n-1)$-manifold $S$
equipped with a triangulation, and to any oriented cobordism $M \maps S
\to S'$ equipped  with a triangulation $\Delta$ compatible with those of
$S$ and $S'$, it assigns an `evolution operator' $Z_0(M)  \maps
Z(S) \to Z(S')$.  We begin by recalling how these are defined, and then
give formula for the evolution operators of the perturbed spin
foam model.   All the formulas are very much like those in the 
previous section.  In fact, partition functions are just the special
case of evolution operators where $S$ and $S'$ are empty.

The Hilbert space $Z(S)$ is defined to have an orthonormal basis given
by spin networks in the dual 2-skeleton of the triangulation of $S$. To
describe the evolution operator $Z_0(M) \maps Z(S) \to Z(S')$, it
thus suffices to give a formula for the transition amplitude between
spin network states $\psi \in Z(S)$, $\psi' \in Z(S')$.  The formula
generalizes that for the partition function of a closed manifold:
\[  \langle \psi', Z_0(M) \psi \rangle  = \]
\[
 \sum_F \prod_{f \in \F } A_0(F,f) \prod_{f \in \F' } A_0(F,f)^{1\over 2} 
 \prod_{e \in \E} A_0(F,e) \prod_{e \in \E'} A_0(F,e)^{1\over 2} 
 \prod_{v \in \V} A_0(F,v).  
\]
Here $\V$ (resp.\ $\E,\F$) denotes the set of vertices (resp.\  edges,
faces) of the triangulation of $M$ that do not intersect the boundary of
$M$, while $\E'$ (resp.\ $\F'$) denotes the set of edges  (resp.\ faces)
that intersect, but do not lie in, the boundary of $M$.  It is important
in this formula to make a consistent choice of the square roots involved.
It then follows that 
\[         Z_0(M)Z_0(M') = Z_0(MM')  \]
whenever $M$ and $M'$ are composable.  

Next we perturb the vertex amplitudes of our spin foam model as in
equation (\ref{perturb}), and define perturbed evolution operators 
$Z(M) \maps Z(S) \to Z(S')$ by 
\[ \langle \psi', Z(M) \psi \rangle  = \]
\[
\sum_F \prod_{f \in \F} A(F,f) \prod_{f \in \F'} A(F,f)^\hf
       \prod_{e \in \E} A(F,e) \prod_{e \in \E'} A(F,e)^\hf
       \prod_{v \in \V} A(F,v).  \]
As before, this has a perturbation expansion 
\be        Z(M) = \sum_{j = 0}^\infty \lambda^j Z_j(M)  \label{z} \ee
with only finitely many nonzero terms.  Moreover, we have   
\[         Z(M)Z(M') = Z(MM')  \]
whenever $M$ and $M'$ are composable.  

To give a formula for $Z_j(M)$, we define a `configuration' as in the
previous section, and for any configuration $(\Delta,C)$ we let the
operator $Z(\Delta,C) \maps Z(S) \to Z(S')$ be given by
\[ \langle \psi', Z(\Delta,C) \psi \rangle  = \]
\[
\sum_F \prod_{f \in \F} A(F,f) \prod_{f \in \F'} A(F,f)^\hf
       \prod_{e \in \E} A(F,e) \prod_{e \in \E'} A(F,e)^\hf
       \prod_{v \notin C} A_0(F,v) \prod_{v \in C} A_1(F,v)  .  \]
Then we have
\be Z_j(M) = \sum_{C \subseteq \V, \; |C| = j} Z(\Delta,C)  \label{zj3} \ee
As before, if our unperturbed spin foam model is invariant under 
the Pachner moves, we can rewrite this as a sum over equivalence
classes of $j$-element configurations.  And as before, the first
few terms of equation (\ref{z}) will be easy to compute in this
case, and should give a good approximation to $Z(M)$ when 
\[         |\lambda|  \ll {1\over |\V|}  . \]

\section{The `Dilute Gas' Limit}  

As a simple application of the ideas above, we now calculate the
evolution operators for a perturbed spin foam model in the $|\V| \to
\infty$ limit, assuming that the original unperturbed spin foam model is
invariant under the Pachner moves.   As we take the limit, we rescale
the coupling constant as follows:
\be             \lambda = {g \over |\V| }    \label{renorm} \ee
for some fixed value of $g$.   This is a simple form of `coupling
constant renormalization', with $g$ playing the role of the renormalized
coupling constant.  Rescaling $\lambda$ this way ensures that the sum
over spin foams is dominated by those where the interaction vertices are
separated from each other.  Physically speaking, this amounts to
considering the limit in which the interaction vertices form a `dilute
gas'.  

In what follows, we fix a connected cobordism $M \maps S \to S'$ and let the
triangulation $\Delta$ of $M$ vary.  We fix constants $k > 0$ and $0 <
\epsilon < 1$, and consider only triangulations for which fewer than $k$
simplices of any dimension share a given 0-simplex, and for which fewer
than $\epsilon |\V|^{1/2}$ of the $n$-simplices intersect the
boundary of $M$.  We claim that with these assumptions, as $|\V| \to
\infty$ the sum (\ref{zj3}) becomes dominated by configurations in a
single equivalence class.

Our argument relies on two plausible conjectures that we have not
proved.   Define a configuration $(\Delta,C)$ to be `separated' if no
two $n$-simplices in $C$ share a 0-simplex, and no $n$-simplex in $C$
intersects the boundary of $M$.   We conjecture that: 
\begin{Alphalist}
\item All separated $j$-element configurations are equivalent.
\item Letting $\Delta$ vary as above and letting $C \subset \V$
be arbitrary, there are only finitely many equivalence 
classes of $j$-element configurations $(\Delta,C)$.  
\end{Alphalist} 
Note that conjecture A depends crucially on the fact that $M$ is
connected, and also the fact that no $n$-simplex in a separated
configuration touches the boundary of $M$.  

Given conjecture A, the value of $Z(\Delta,C)$ is the same for every 
separated $j$-element configuration $(\Delta,C)$.  Call this value
$\alpha_j(M)$.   Then we claim that as $|\V| \to \infty$, we have the 
asymptotic formula 
\be  Z_j(M) \sim {|\V|^j \over j!}  \alpha_j(M) \label{asymp} .\ee
In other words, separated configurations dominate the sum over spin
foams.

To prove equation (\ref{asymp}), note that in equation (\ref{zj3}) we
are computing $Z_j(M)$ as a sum over all $j$-element configurations with
the same underlying triangulation.   There are exactly ${|\V| \choose
j}$ such configurations.  In the limit as $|\V| \to \infty$, this number
is asymptotic to $|\V|^j / j!$.  Of these configurations, fewer than 
\[     (k (j-1) + \epsilon |\V|^{1/2}) {|\V| \choose j-1}  \]
are nonseparated, since to form a nonseparated configuration we can
choose the first $j-1$ of the $n$-simplices arbitrarily, but the last
one must either share at least one 0-simplex with the rest, or intersect
the boundary.  As $|\V| \to \infty$, this number is on the order of
$|\V|^{j - 1/2}$, so in the $|\V| \to \infty$ limit almost all the
configurations being summed over are separated.  Each such configuration
contributes $\alpha_j(M)$ to the sum for $Z_j(M)$, so the total
contribution of the separated configurations is asymptotic to
$\alpha_j(M) |\V|^j / j!$.  Thus to finish proving equation
(\ref{asymp}) it suffices to check that the nonseparated configurations
contribute an amount of order at most $|\V|^{j - 1/2}$ to the sum for
$Z_j(M)$.    Because there are on the order of $|\V|^{j - 1/2}$, and two
configurations in the same equivalence contribute the same amount, this
follows from conjecture B.

Equations (\ref{renorm}) and (\ref{asymp}) imply that 
\be \lim_{|\V| \to \infty} \lambda^j Z_j(M) = {g^j \over j!} \alpha_j(M) .
\label{limit} \ee
This means that if we renormalize the coupling constant appropriately
while taking the $|\V| \to \infty$ limit, each term in the power series
expansion for the evolution operator $Z(M)$ converges to a quantity that
can be explicitly computed using any triangulation of $M$ that is large
enough to contain a separated $j$-element configuration.  

Naively one might wish to conclude from equation (\ref{limit}) that
\[ 
\lim_{|\V| \to \infty} Z(M) = \sum_{j=0}^\infty {g^j \over j!} \alpha_j(M),
\]
but of course there are two problems.  First, the power series on the
right-hand side of this equation might not converge.  Second, the
fact that one power series converges to another term-by-term does not
imply that the sum of the first converges to the sum of the second. 
In what follows we shall treat the first problem but not the second.
In other words, we show that for any choice of the interaction amplitude
$A_1(F,v)$, the power series 
\be     
Z_g(M) = \sum_{j=0}^\infty {g^j \over j!} \alpha_j(M) .\label{zg} 
\ee
converges for all values of $g$.  

When $M,M'$ are connected cobordisms and $MM'$ is their composite, we
have 
\[     \alpha_j(M)\alpha_k(M') = \alpha_{j+k}(MM').  \]
This implies, at the level of formal power series, that
\be  Z_g(M) Z_{g'}(M') = Z_{g+g'}(MM') . \label{prod} \ee
A particularly important special case of this equation occurs
when $M \maps S \to S$ is the cylinder $S \times I$, where $I$ is the
closed unit interval.   In this case we have $(S \times I)^2 = 
S \times I$, so 
\[  Z_g(M)Z_{g'}(S \times I) =  Z_{g+g'}(S \times I) . \] 
This implies that 
\ba    Z_g(S \times I) &=& \exp(-gH_S) Z_0(S \times I)  \label{exp} \\
                       &=& Z_0(S \times I) \exp(-gH_S) \nonumber \ea
for some operator $H_S$, which is given explicitly by 
\[     H_S = -\alpha_1(S \times I) .\] 

We are now in a position to see that the power series for
$Z_g(M)$ in equation (\ref{zg}) converges for all values of $g$. 
When $M$ is the cylinder $S \times I$, the power series for $Z_g(M)$ 
converges by equation (\ref{exp}), because the power series for 
$\exp(-g H_S)$ converges.  In general, we can write any connected 
cobordism $M$ as a composite $M_1 M_2 M_3$ where $M_2$ is a cylinder, and 
use equation (\ref{prod}) to write $Z_g(M)$ as $Z_0(M_1)Z_g(M_1)Z_0(M_2)$, 
thus reducing to the cylinder case.  For the sake of completeness, it is 
also useful to define $Z_g(M)$ when $M$ is not connected.  In this case, 
we define $Z_g(M)$ to be the (tensor) product of evolution operators for its
connected components.  This again allows us to write $Z_g(M)$ as a
convergent power series in $g$.  

It is clear that the perturbative spin foam models under
consideration are not topological quantum field theories.   In a
topological theory, the evolution operator associated to a cylinder is 
just the projection onto the subspace of physical states.  Here,
however, we have 1-parameter semigroup of evolution operators depending
on $g$.   At $g = 0$, this reduces to the projection operator for
the topological quantum field theory being perturbed about.

It may seem puzzling that the parameter $g$, originally introduced as a
renormalized coupling constant, is now playing a role a bit like `time'.
In fact, we should not think of $g$ as `time', but as an extensive
parameter measuring the mean number of interaction vertices.   The
reason is that in formula (\ref{zg}) for the evolution operator, the
term corresponding to a spin foam with $j$ interaction vertices is
weighted by a factor of $g^j/j!$.  Up to a normalization factor, this is
just the Poisson distribution with mean $g$.   (To normalize this
Poisson distribution, we could divide $Z_g(M)$ by $\exp(g)$.)  We can thus
think of $Z_g(M)$ as the evolution operator for a spacetime manifold $M$
containing a large spin foam for which the interaction vertices
are separated and their number is distributed according to the Poisson
distribution with mean $g$.  

This interpretation of $g$ as an extensive parameter is clearest in the
case of 2-dimensional Yang-Mills theory.  Witten \cite{W} showed that 2d
Yang-Mills theory can be described as a perturbation of 2d $BF$ theory,
and his work fits nicely into the language of spin foam models
\cite{FK}.  In this case, the renormalized coupling constant $g$
turns out to have the physical significance of {\it area}.  To illustrate
the results above, we now turn to this example.

Let $R$ be a complete set of inequivalent irreducible representations of
a compact Lie group $G$.  Of course, $R$ is finite only when $G$ is, so
our results strictly apply only to that case, but apart from some 
convergence issues that we mention below, they generalize
straightforwardly to the compact case.  To simplify our description of
the face, edge, and vertex amplitudes of 2d $BF$ theory, we can use the
orientation of $M$ to consistently orient all the faces.  The amplitude
for  a face labelled by the representation $\rho$ is just $\dim(\rho)$. 
Each edge has two faces incident to it, which must be labelled with the
same representation $\rho$ for there to be an intertwiner labelling the
edge; the amplitude for the edge is then $\dim(\rho)^{-1}$.   Finally,
each vertex has a collection of faces incident to it, which must all be
labelled by the same representation $\rho$, and the amplitude for the
vertex is then $\dim(\rho)$.   

It follows that when $M$ is a connected surface without boundary, its
partition function is  
 \[      Z_0(M) = \sum_{\rho \in R} \dim(\rho)^{\chi(M)} .  \]   
(This may not converge when $G$ is not finite.) We can consider an
arbitrary perturbation of the vertex amplitudes, which is necessarily of
the form 
\[       A_1(F,v) = f(\rho)  \] 
where $f \maps R \to \C$ and $\rho$ is the representation labelling all
the faces incident to $v$.  If we take $S$ to be the circle triangulated
using a single 1-simplex, then $Z(S)$ has an orthonormal basis
$\psi_\rho$ with one element for each irreducible representation $\rho
\in R$, and the operator $H_S$ given in equation (\ref{exp}) has the
form 
\[   H_S \psi_\rho = -f(\rho) \psi_\rho  . \] 
It follows that  
\[ Z_g(M) = \sum_{\rho \in R} \dim(\rho)^{\chi(M)} \, e^{g f(\rho)}.  \] 
This may converge even when $Z_0(M)$ does not.  In particular, if the
Lie algebra of $G$ is equipped with an invariant inner product, and we
let $f(\rho)$ be the minus the Casimir of the representation $\rho$, the
sum for $Z_g(M)$ converges whenever $g > 0$.  In this case $H_S$ is the
Hamiltonian for 2d Yang-Mills theory, and $Z_g(M)$ is the partition
function for 2d Euclidean Yang-Mills theory on a surface with area $g$.
Taking $g$ imaginary gives the Lorentzian theory.

Besides Yang-Mills theory, this setup gives many other 
triangulation-independent spin foam models in 2 dimensions.  If we take
$G$ to be finite, any function $f \maps R \to \C$ will  give such a
model.  The most interesting models are the `unitary' ones, where $f$ is
real-valued.   If $G$ is compact, we obtain a unitary model with
convergent Euclidean partition functions for surfaces with high genus
from any function $f \maps R \to \R$ that grows sufficiently fast.  
We can also work with a quantum group at root of unity instead of
a group.  This gives theories that are perturbations of the $G/G$
gauged WZW model instead of $BF$ theory.
 
Since the dilute gas limit gives a host of 
triangulation-independent spin foam models in 2 dimensions, we might
hope for similar results in higher dimensions.  However, these hopes
are in vain.  To see this, let $Z_0(S)$ be the Hilbert space of
physical states associated to the manifold $S$, i.e., the range of the
projection $Z_0(S \times I)$.   Since $H_S = H_S Z_0(S \times I)$, 
$H_S$ is completely determined by its value on physical states.   In
what follows, we show that on physical states, $H_S$ is just a multiple
of the identity in theories for which $Z_0(S^{n-1})$ is 1-dimensional. 
This includes all the theories of Turaev-Viro type in 3 dimensions and
of Crane-Yetter type in 4 dimensions.  Thus for such theories, any
perturbation of the vertex amplitudes has a trivial effect in the dilute
gas limit: for any cobordism $M$, we have
\[        Z_g(M) = \exp(-cg) Z_0(M) \]
for some constant $c$.  

To see that $H_S$ is a multiple of the identity on physical states when
$Z_0(S^{n-1})$ is 1-dimensional, we use some facts about the structure
of topological quantum field theories \cite{Sawin}.   First, the obvious
cobordism from $S^{n-1} \cup S^{n-1}$ to  $S^{n-1}$ makes $A =
Z_0(S^{n-1})$ into a commutative algebra.   Second, for every compact
connected oriented $(n-1)$-manifold $S$, the obvious cobordism
\[     M \maps S^{n-1} \cup S \to S   \]
makes $Z_0(S)$ into an $A$-module.   The cobordism $D^n \maps \emptyset
\to S^{n-1}$ gives an operator
\[       \alpha_1(D^n) \maps \C \to A , \]
and the image of $1 \in \C$ under this operator gives a special element of
$A$ which we call $-H$.  This has the property that for any physical
state $\psi \in Z_0(S)$, 
\[    H_S \psi = H \psi ,\]
where the right-hand side is defined using the action of $H \in A$ on
$Z_0(S)$.  In Figure 2 we show this action of $H$ on $Z_0(S)$;
here we have omitted all the $n$-simplices in the triangulation
except the one corresponding to the single interaction vertex, shown in 
black.  If $Z_0(S)$ is 1-dimensional, $H$ is a multiple of the
identity in $A$, so $H_S$ is a fixed multiple of the identity,
independent of $S$.  

\vbox{
\vskip 1em
\centerline{\epsfysize=2.0in\epsfbox{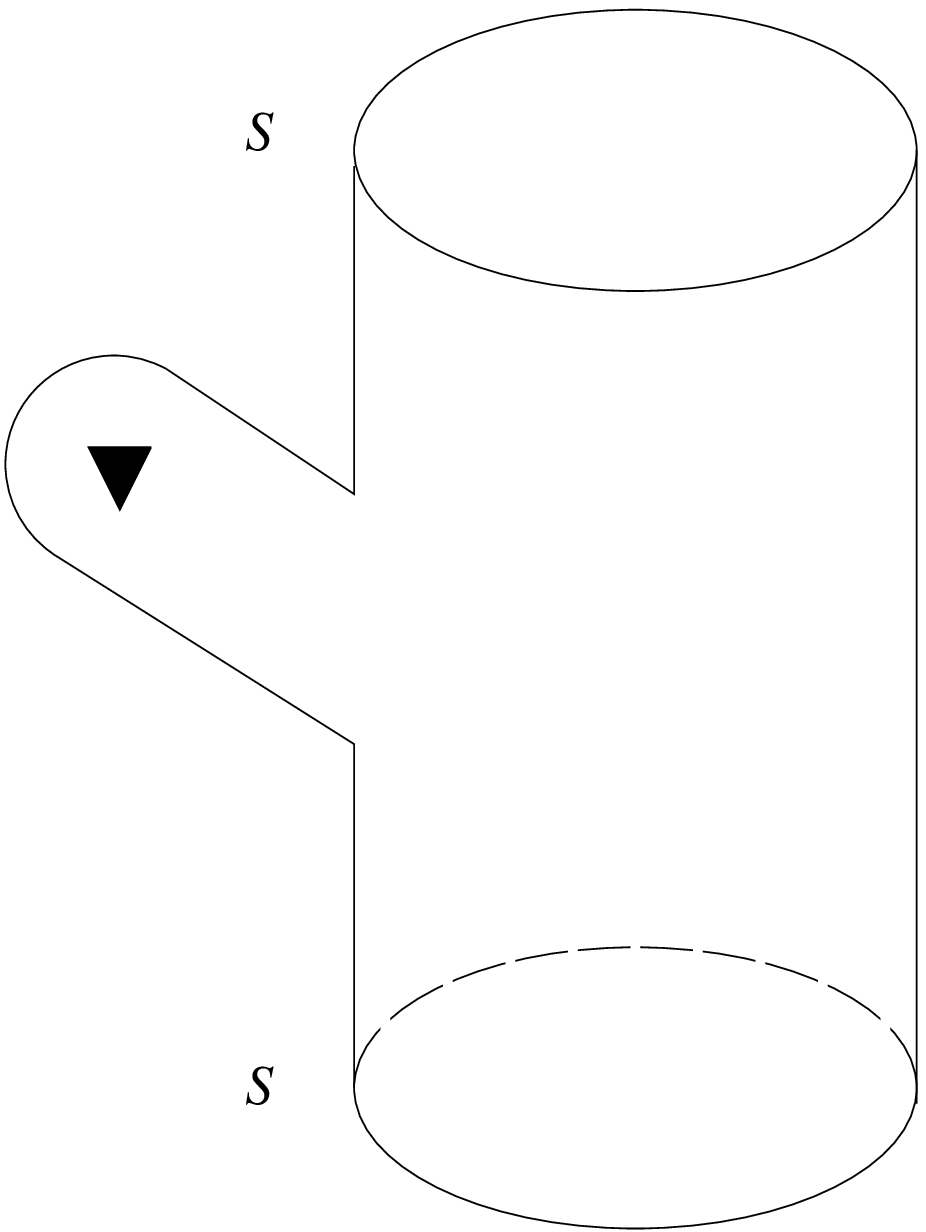}} \medskip
\centerline{2.  The action of $H$ on physical states}
\medskip
}

\section{Conclusions}

While the spin foam perturbation theory developed here is applicable  to
a wide variety of situations, we have seen that one simple-minded  way
of using it to construct triangulation-independent spin foam models 
gives trivial results except when spacetime has dimension $2$.  It is
worth reflecting on exactly why this happens.  Technically, the reason
is that in higher dimensions the TQFTs being perturbed about assign a 
1-dimensional physical Hilbert space to the sphere.  The reason for
this, in turn, is that these theories are obtained by quantizing
theories of flat connections, and the moduli space of flat connections
on $S^{n-1}$ is just a point when $n \ne 2$.   

A perhaps more illuminating explanation is that in the dilute gas limit, 
nontrivial interactions occur at a discrete set of points.  Removing a 
point from a manifold changes its moduli space of flat connections only 
in the 2-dimensional case, because only then is there a noncontractible
loop going around the removed point.   Thus, only in this case
can the operator shown in Figure 2 have an interesting effect.  

This suggests a number of ideas.  Naively, one might hope that 
perturbing edge or face amplitudes could give more interesting 
results in higher dimensions.  However, it does not.  Everything we have 
done for perturbations of vertex amplitudes works very similarly
for perturbations of edge and face amplitudes.    A more interesting
idea is to consider more general state sum models in which all
the cells of the dual skeleton are labelled, not just those of
dimensions 0 and 1.  Such models correspond to field theories based
not on connections, but on categorified analogues of connections,
such as connective structures on gerbs \cite{Bryl}.   Some 
topological quantum field theories of this type have been studied 
by Yetter \cite{Y}, Porter \cite{P2}, and Mackaay \cite{Mackaay,Mackaay2}.
In the dilute gas limit, perturbations of these should give 
triangulation-independent models that are not topological quantum 
field theories.   These should be more interesting in dimensions
above 2 than the examples considered here.  

Ultimately, to make contact with real physics, one must study 
theories with propagating degrees of freedom.   Whether perturbation
about a topological quantum field theory can really be useful here 
is not clear.  However, for it to have any chance of being useful, 
we must go beyond the dilute gas limit, and explore the effect of
nonseparated configurations in the perturbation expansion.  

\subsection*{Acknowledgements}  I thank Laurent
Freidel, Kirill Krasnov, Carlo Rovelli, Stephen Sawin, and Lee
Smolin for helpful conversations and correspondence.

\end{document}